\documentclass[runningheads]{llncs}
\usepackage{graphicx}
\usepackage{subcaption}
\usepackage{amssymb}
\usepackage{amsfonts}
\usepackage{amsmath}
\usepackage{url}
\usepackage{float}
\usepackage{multirow}
\usepackage{bm}

\usepackage{pifont}% http://ctan.org/pkg/pifont
\newcommand{\cmark}{\ding{51}}%
\newcommand{\xmark}{\ding{55}}%

\DeclareMathOperator*{\argmin}{\arg\min}   % rbp
\DeclareMathOperator*{\argmax}{\arg\max}   % rbp
\begin{document}
\raggedbottom

\title{Mask2Lesion: Mask-Constrained Adversarial Skin Lesion Image Synthesis}
\titlerunning{Mask2Lesion: Mask-Constrained Adversarial Skin Lesion Image  Synthesis}
% \author{Kumar Abhishek\orcidID{301371793}}
\author{Kumar Abhishek and Ghassan Hamarneh}
\authorrunning{Abhishek and Hamarneh}
\institute{School of Computing Science, Simon Fraser University, Canada
\email{\{kabhishe,hamarneh\}@sfu.ca}}
\maketitle              % typeset the header of the contribution
\begin{abstract}
Skin lesion segmentation is a vital task in skin cancer diagnosis and further treatment. Although deep learning based approaches have significantly improved the segmentation accuracy, these algorithms are still reliant on having a large enough dataset in order to achieve adequate results. Inspired by the immense success of generative adversarial networks (GANs), we propose a GAN-based augmentation of the original dataset in order to improve the segmentation performance. In particular, we use the segmentation masks available in the training dataset to train the Mask2Lesion model, and use the model to generate new lesion images given any arbitrary mask, which are then used to augment the original training dataset. We test Mask2Lesion augmentation on the ISBI ISIC 2017 Skin Lesion Segmentation Challenge dataset and achieve an improvement of $5.17\%$ in the mean Dice score as compared to a model trained with only classical data augmentation techniques.

\end{abstract}
\section{Introduction}

Melanoma, a type of skin cancer, although represents a small fraction of all skin cancers in the USA, accounts for over 75\% of all skin cancer related fatalities~\cite{rogers2015incidence}, and is responsible for over 10,000 deaths annually across the country~\cite{SkinCancer}. However, studies have shown that the survival rates of patients improve drastically with early diagnosis. Efficient assessment of dermoscopic images for indicators of melanoma is an important component of early diagnosis and improved patient prognosis. Automated methods to extract image features indicative of skin lesions are promising tools for dermatologists. Based on core methods such as the 7-point checklist~\cite{7Point}, the ABCD (Asymmetry, Border, Color, and Differential structure) rule~\cite{ABCD}, and the CASH (Color, Architecture, Symmetry, and Homogeneity) algorithm~\cite{CASH}, deep learning methods can aid the diagnosis of skin lesion images. However, these methods use hand-crafted features, and therefore rely on an accurate segmentation of the lesion~\cite{barata2018survey}. Moreover, lesion segmentations have been used to assist melanoma diagnosis~\cite{hosny2019classification,sumithra2015segmentation,yan2019melanoma}. This motivates the use of deep learning based computer-aided diagnosis systems to improve the accuracy and sensitivity of melanoma detection methods.

Recent works on skin lesion segmentation using deep learning have shown significant improvements in segmentation accuracy. Yuan et al.~\cite{Yuan} used a 19-layer deep fully convolutional network with a Jaccard distance based loss function that is trained end-to-end to segment skin lesions. Mirikharaji et al.~\cite{mirikharaji2018deep} proposed a deep auto-context architecture to use image appearance information along with the contextual information to improve segmentation results. Yu et al.~\cite{yu2017automated} proposed using a deep residual network architecture with several blocks stacked together to improve the representative capability of the network and therefore increase the segmentation accuracy.

Generative adversarial networks (GANs), proposed by Goodfellow et al.~\cite{GAN} have been immensely popular in realistic image generation tasks.  Numerous variations of these generative models have been developed for a variety of applications, including text to image synthesis and video generation~\cite{ReedAYLSL16,VondrickPT16}. GANs have also been used to generate various medical imaging modalities, such as generating liver lesion images to augment the CT lesion classification training dataset~\cite{Liver}, generating chest X-ray images to augment the dataset for abnormality detection~\cite{madani2018chest}, and generating brain CT images from corresponding brain MR images~\cite{wolterink2017deep}. Skin lesion synthesis tasks have also relied upon GAN-based approaches, such as generating images of benign and malignant skin lesions~\cite{baur2018generating}, modeling skin lesions using semantic label maps and superpixels in order to generate new lesion images~\cite{bissoto2018skin}, and generating skin lesions along with their corresponding segmentation masks~\cite{Pollastri2019}.

In this work, we propose to use lesion masks to generate synthetic lesion images in order to augment the segmentation training dataset and improve skin lesion segmentation performance. Isola et al.~\cite{pix2pix} and Zhu et al.~\cite{CycleGAN} have shown that it is possible to generate high resolution realistic images from object boundaries. An inherent advantage of using lesion masks to generate skin lesion images is that the newly generated images can be used for training the segmentation network without needing to be annotated. To the best of our knowledge, this is the first work towards generating images for medical applications from shape. In particular, we synthesize skin lesion images from lesion masks.

The paper is structured as follows: we discuss the proposed approach in Section~\ref{sec:Method}, describe the dataset and experimental details in Section~\ref{sec:data}, and analyze the quantitative and qualitative results of our proposed approach in Section~\ref{sec:Results}. Section~\ref{sec:Conclusion} concludes the paper.

\section{Method}    \label{sec:Method}

\subsection{Method Overview}
The purpose of our method is to synthesize segmentation training data which is then used to augment the existing data for training a segmentation network. We model this as an image-to-image translation task where we train a deep neural network model, called Mask2Lesion, to generate the synthetic data. In particular, we translate images containing binary segmentation masks, which highlight the area of a target skin lesion, to a skin image containing a lesion confined to that binary mask, making it a paired image-to-image translation task. To this end, we train a network with skin lesion images and their corresponding masks. Such training data is also typically provided for training segmentation methods. Our deep network is based on the pix2pix conditional generative adversarial network (GAN), described in Section~\ref{subsec:I2I}. With the ability to translate a binary mask to a corresponding image containing a lesion delineated by the mask, we can then turn our attention to creating synthesized masks (via different approaches), and rely on our trained Mask2Lesion model to generate the corresponding images. Given a training dataset of images and segmentation masks, with or without augmentation (performed using Mask2Lesion or otherwise), we can then train a segmentation network. The segmentation network used here is described in Section~\ref{sec:data}.

\subsection{Segmentation Masks}
\begin{figure}[H]
  \centering
  \includegraphics[width=0.65\textwidth]{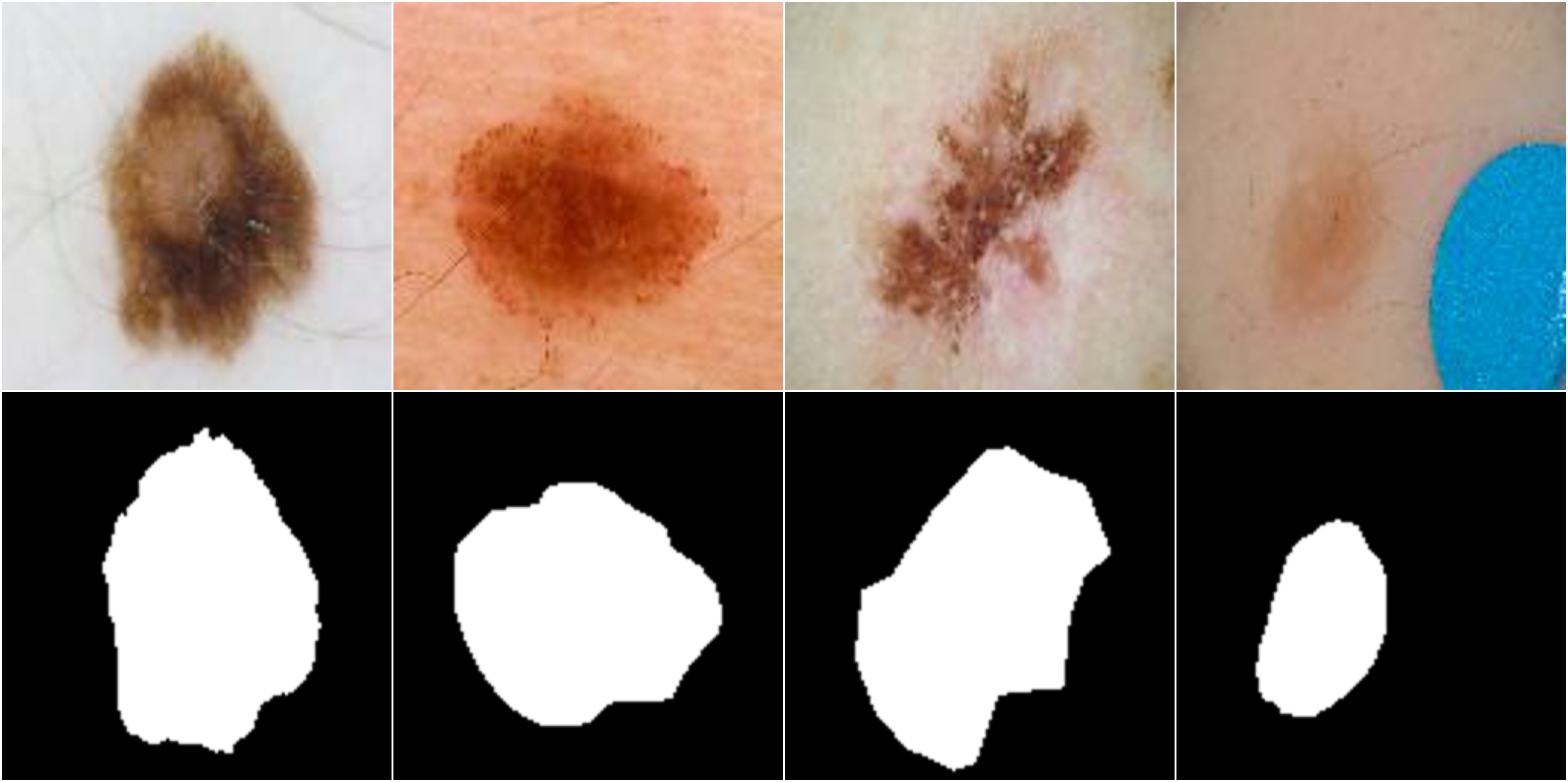}
  \caption{A few sample images from the ISIC training dataset along with the corresponding segmentation masks. Note the presence of artifacts in some of the images.}
  \label{fig:SampleMask}
\end{figure}

We propose to use lesion segmentation masks as input to the generative algorithm, making it easy to produce a large number of inputs. Since the ISIC 2017 Skin Lesion Segmentation Challenge dataset~\cite{ISIC2017} used for the segmentation task has ground truth segmentation masks available, they can be used as inputs to the generative algorithm to synthesize skin lesions, thus creating new pairs of lesion images and their masks. Figure~\ref{fig:SampleMask} shows four sample lesion images with their corresponding segmentation masks.

\subsection{Image-to-Image Translation Network} \label{subsec:I2I}

The paired image-to-image translation model proposed by Isola et al.~\cite{pix2pix} uses a conditional GAN to generate images. Unlike traditional GANs which learn a mapping from a random noise vector to an output image, conditional GANs learn a mapping from an observed image $x$ and a random noise vector $z$ to an output image $y$. The two components of a conditional GAN are a generator and a discriminator. The generator G is trained to produce output images, $G \ : \ \{x, z\} \to y$ which are ``realistic", meaning they cannot be distinguished from the original images. The discriminator D tries to distinguish between the original images and the output of the generator G. The two components can be estimated using deep neural networks. This conditional GAN is trained in an adversarial manner, and the objective function can be written as
\begin{equation}
    \mathcal{L}_{cGAN} (G, D) = \mathbb{E}_{x,y} \left[\text{log} \ D(x, y)\right] \ + \ \mathbb{E}_{x,z} \left[\text{log} \ \left(1 - D(x, G(x,z))\right)\right],
\end{equation}
where the generator $G$ tries to minimize this objective function and the discriminator $D$ tries to maximize it. The optimal solution is obtained using this minimax game
\begin{equation}
    G^\textbf{*} = \argmin_G \ \argmax_D \ \mathcal{L}_{cGAN} (G, D).
\end{equation}
This is different from an unconditional GAN where the discriminator $D$ does not observe the input image $x$.

\subsubsection{Generator Architecture:}
Since the output of the generator shares the underlying structure with the input, an encoder decoder architecture with skip connections has been chosen as the generator. We use U-Net~\cite{U-Net} with an L1 loss because in its attempt to fool the discriminator, L2 loss tends to produce more blurry generator outputs. The U-Net has a fully convolutional neural network architecture consisting of two paths - a contracting path and a symmetric expansive path. Skip-connections containing feature maps from the contracting path to the symmetrically corresponding layer's upsampled feature maps in the expanding path assist recovery of the full spatial resolution at the network output~\cite{li2018visualizing}.

\subsubsection{Discriminator Architecture:}
While using the L1 loss for the generator ensures that the low frequency details are accurately captured, it is also important to model the high-frequency structure of the image. This is achieved by using a PatchGAN~\cite{pix2pix}, a discriminator architecture which penalizes structure at local image patch level. As a result, the image is divided into several (overlapping) patches, each of which is labeled by the discriminator as `real' or `fake', and the overall output of the discriminator is the average of the individual responses.

\begin{figure}[H]
  \centering
  \includegraphics[width=0.6\textwidth]{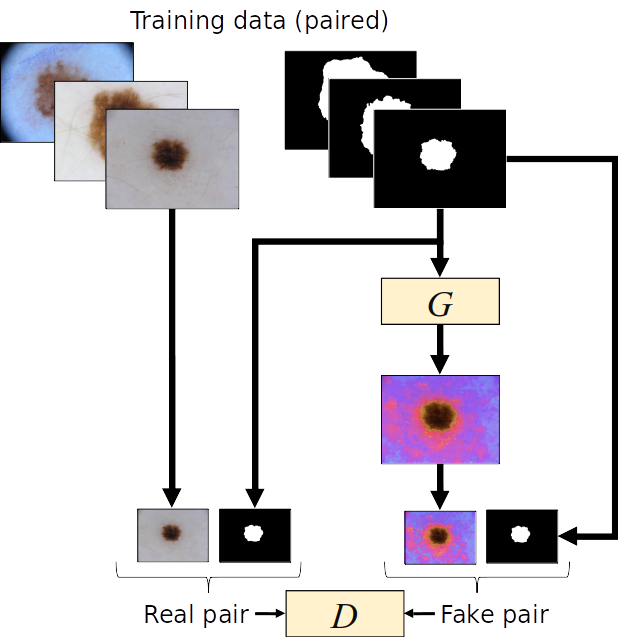}
  \caption{The proposed Mask2Lesion algorithm.}
  \label{fig:approach}
\end{figure}

\smallskip
Figure~\ref{fig:approach} shows a high level overview of the Mask2Lesion algorithm.

\section{Data and Experimental Details} \label{sec:data}
The dataset used for evaluation of the proposed approach was obtained from the $2017$ ISBI ISIC Skin Lesion Analysis Towards Melanoma Detection: Lesion Segmentation Challenge~\cite{ISIC2017}, and contains $2000$ training images and $150$ test images. All the images and their corresponding ground truth segmentation masks were resized to $128\times128$ pixels using nearest neighbor interpolation from the SciPy library. 

The Mask2Lesion model was trained for 200 epochs. For both the generator and the discriminator, all convolution operations used $4\times4$ spatial filters with a stride of $2$. Each convolution layer (except the first) consists of convolution, batch normalization, dropout (with a keep probability of $0.5$), and ReLU activation. The encoder (the contracting path of the U-Net) uses leaky ReLUs with a slope of $0.2$, while the decoder (the expansive path) uses ReLUs. For the PatchGAN, a $70\times70$ patch is processed from the input image, which assigns a score to a $30\times30$ patch of the image.

As the goal of this work is to demonstrate the efficacy of the proposed Mask2Lesion model in augmenting the dataset for segmentation, we use U-Net~\cite{U-Net} as a baseline segmentation network, and optimize it with mini-batch stochastic gradient descent (SGD) with a batch size of 32. In order to evaluate the segmentation performance with and without GAN based augmentation, we train and evaluate four segmentation networks, and we use the following abbreviations to denote them while reporting results: (i) \textbf{NoAug}: trained on only the original training dataset without any augmentation, (ii) \textbf{ClassicAug}: trained on the original training dataset augmented with classical augmentation techniques (rotation, flipping, etc.), (iii) \textbf{Mask2LesionAug}: trained on the original training dataset augmented with Mask2Lesion outputs on masks from the training dataset, and (iv) \textbf{AllAug}: trained on the original dataset augmented with classical augmentation as well as Mask2Lesion outputs on masks from the training dataset. For all the segmentation networks, we report the metrics used in the challenge~\cite{ISIC2017} - Dice coefficient, sensitivity, specificity, and pixel-wise accuracy.

\begin{figure}[H]
\centering
\includegraphics[width=0.999\textwidth]{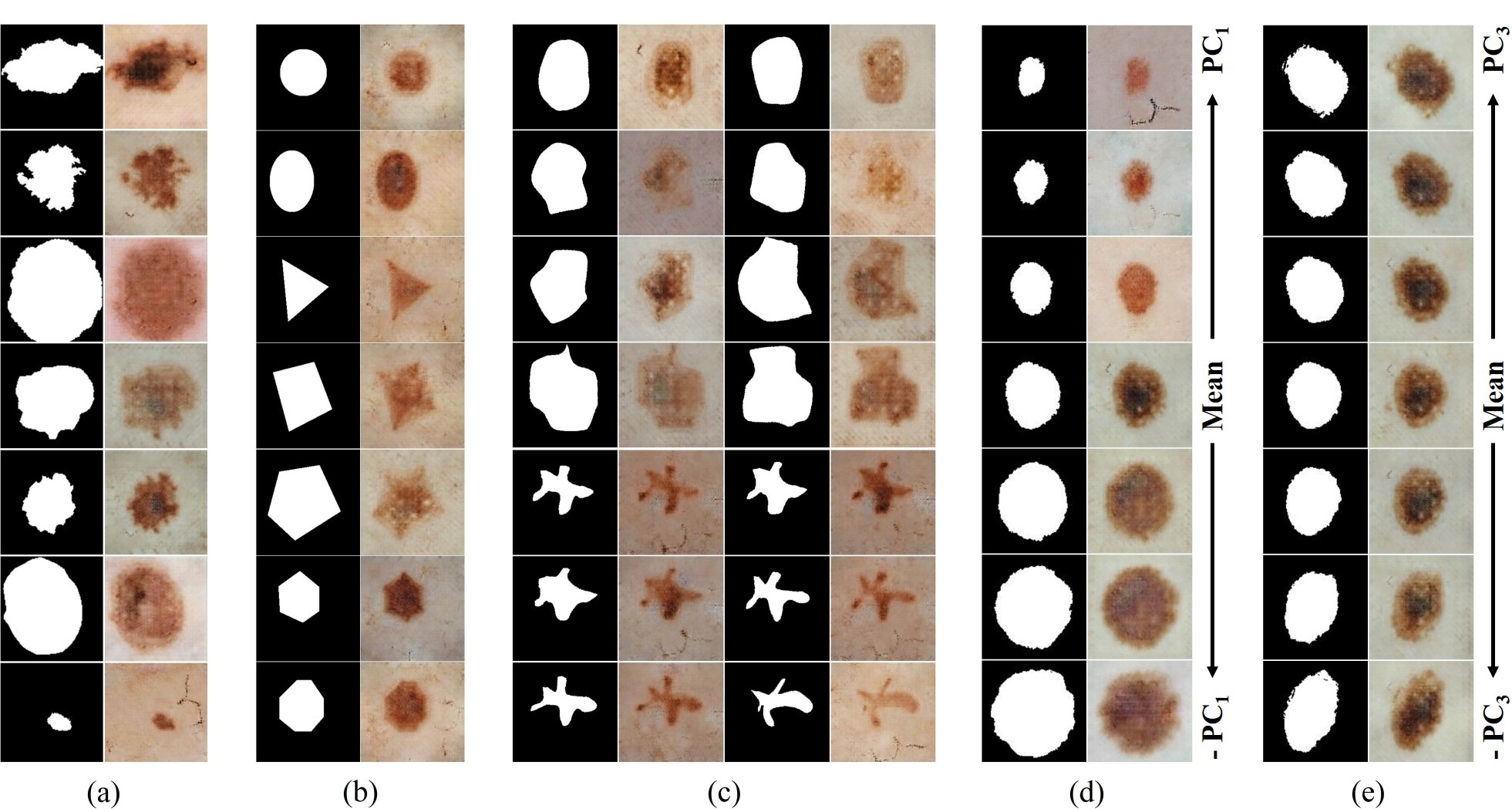}
\caption{(a) Segmentation masks from the ISIC dataset fed to Mask2Lesion and the corresponding generated lesion images. (b) Simple geometric shapes as masks and the corresponding outputs. (c) Elastic deformations applied to hand drawn masks using DeformIt and the corresponding synthesized lesion images. (d),(e) PCA-based deformations applied to segmentation masks and the corresponding Mask2Lesion outputs.}
\label{fig:SegMask_p2p}
\end{figure}

\section{Results and Discussion}    \label{sec:Results}
We use the segmentation masks from the ISIC dataset as inputs to the Mask2Les- %manually broken word
ion model, and the corresponding generated lesion images are shown in Figure~\ref{fig:SegMask_p2p}(a). We see that the synthesized lesions express variance in appearances and textures. 

Next, we test Mask2Lesion by using simple geometric shapes as masks, showing that synthesized images are well constrained by the mask boundaries (Figure~\ref{fig:SegMask_p2p}(b)). We also test the adaptability of Mask2Lesion to hand-drawn masks. We draw two shapes - a large blob and a star shape, and then apply varying degrees of elastic deformations to them using DeformIt~\cite{hamarneh2008simulation}. These masks are then used as inputs to the Mask2Lesion model and the corresponding outputs are shown in Figure~\ref{fig:SegMask_p2p}(c). Furthermore, we apply deformations using a PCA-based shape model to segmentation masks. In particular, we generate new masks by weighting the first and the third principal components in the range $[-1, 1]$ in order to incorporate size and orientation changes (Figures~\ref{fig:SegMask_p2p}(d) and (e) respectively), and use these to generate lesion images. We note that the goal for testing on geometric shapes, hand-drawn masks, and masks deformed using PCA-based shape modeling is to showcase our method's ability to generate skin lesion images confined to the user-specified input masks, regardless of their complexity.

\begin{figure}[H]
  \centering
  \includegraphics[width=0.8\textwidth]{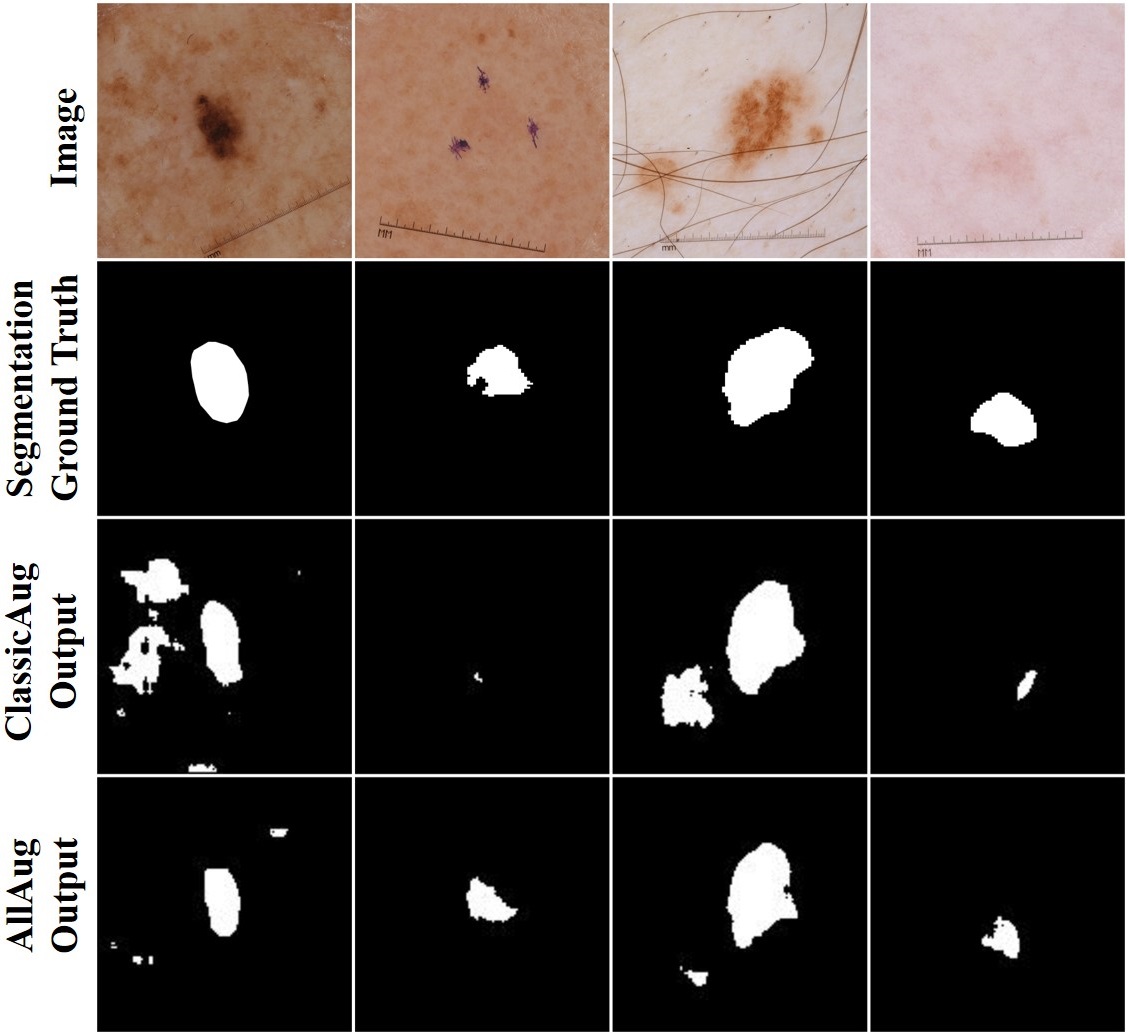}
  \caption{Improved segmentation accuracy with AllAug. The first row shows the test image samples, the second row shows the segmentation ground truths and the third and the fourth rows show the segmentations obtained from the ClassicAug and AllAug respectively.}
  \label{fig:GoodSeg}
\end{figure}

Table~\ref{tab:my-table1} shows the quantitative results for the test images evaluated using the four trained segmentation networks. We see that Mask2LesionAug outperforms ClassicAug in Dice coefficient, sensitivity, and specificity. Moreover, AllAug (which combines both classical as well as Mask2Lesion-based augmentation) outperforms ClassicAug in all four metrics, and achieves a $5.17\%$ improvement in the mean Dice coefficient. Figure~\ref{fig:GoodSeg} shows samples from the test dataset for which the segmentation accuracy significantly improved with AllAug. The outputs of AllAug are much more closer to the respective ground truths and have fewer false positives as compared to ClassicAug.

\begin{table}[]
\centering
\scriptsize
\caption{Quantitative results for segmentation (Mean $\pm$ standard error)}
\label{tab:my-table1}
{\renewcommand{\arraystretch}{1.5}
\begin{tabular}{rc|c|c|c|c|c}
\hline
\multicolumn{3}{c|}{\textbf{Method}}                & NoAug                            & ClassicAug                       & Mask2LesionAug                   & AllAug                           \\ \hline
\multirow{2}{*}{\begin{tabular}[c]{@{}c@{}}\textbf{Aug.}\\\textbf{Method}\end{tabular}} & \multirow{2}{*}{} & Classical   & \xmark            & \cmark            & \xmark            & \cmark            \\ \cline{3-7} 
                                                                              &                   & Mask2Lesion & \xmark            & \xmark           & \cmark            & \cmark            \\ \hline
\multicolumn{3}{c|}{\textbf{Dice}}                  & $0.7723 \pm 0.0185$ & $0.7743 \pm 0.0203$ & $0.7849 \pm 0.0160$ & $\bm{0.8144} \pm \bm{0.0160}$ \\ \hline
\multicolumn{3}{c|}{\textbf{Accuracy}}              & $0.9316 \pm 0.0089$ & $0.9321 \pm 0.0086$ & $0.9311 \pm 0.0087$ & $\bm{0.9375} \pm \bm{0.0091}$ \\ \hline
\multicolumn{3}{c|}{\textbf{Sensitivity}}           & $0.7798 \pm 0.0211$ & $0.8094 \pm 0.0222$ & $0.8197 \pm 0.0186$ & $\bm{0.8197} \pm \bm{0.0182}$ \\ \hline
\multicolumn{3}{c|}{\textbf{Specificity}}           & $0.9744 \pm 0.0035$ & $0.9672 \pm 0.0047$ & $0.9698 \pm 0.0045$ & $\bm{0.9762} \pm \bm{0.0038}$ \\ \hline
\end{tabular}}
\end{table}

\begin{figure*}[]
  \centering
  \includegraphics[width=0.32\textwidth]{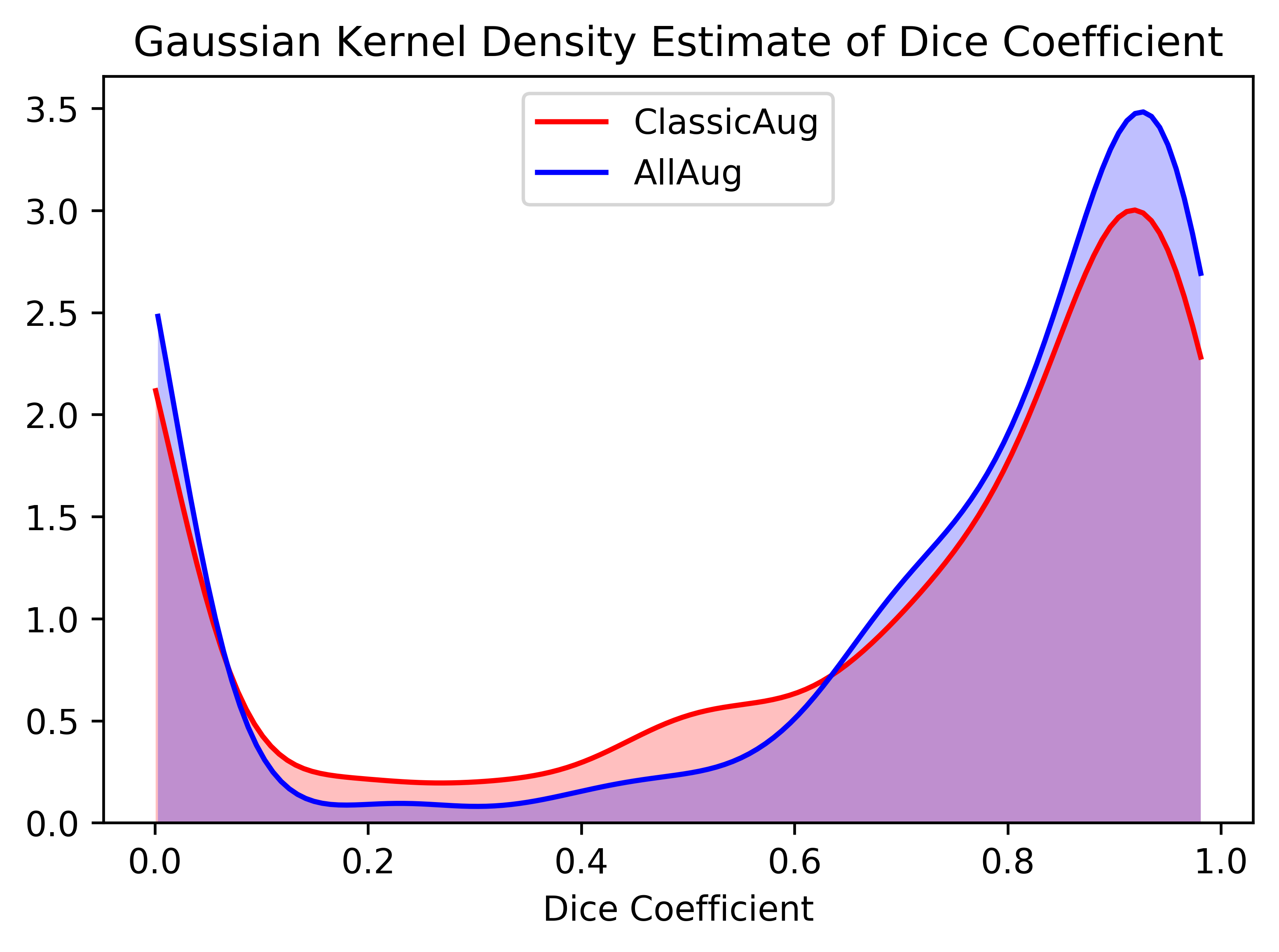}
  \includegraphics[width=0.32\textwidth]{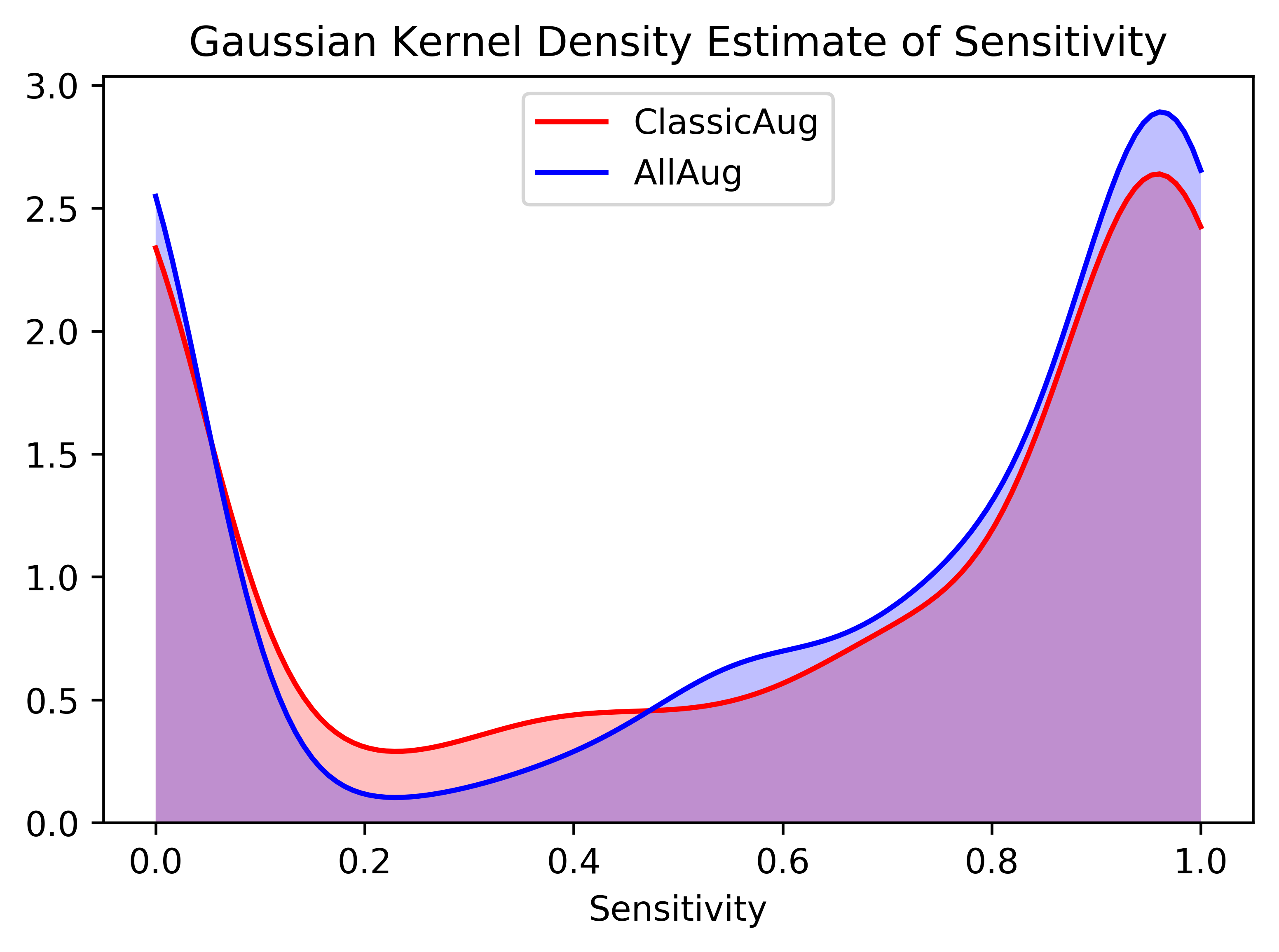}
  \includegraphics[width=0.32\textwidth]{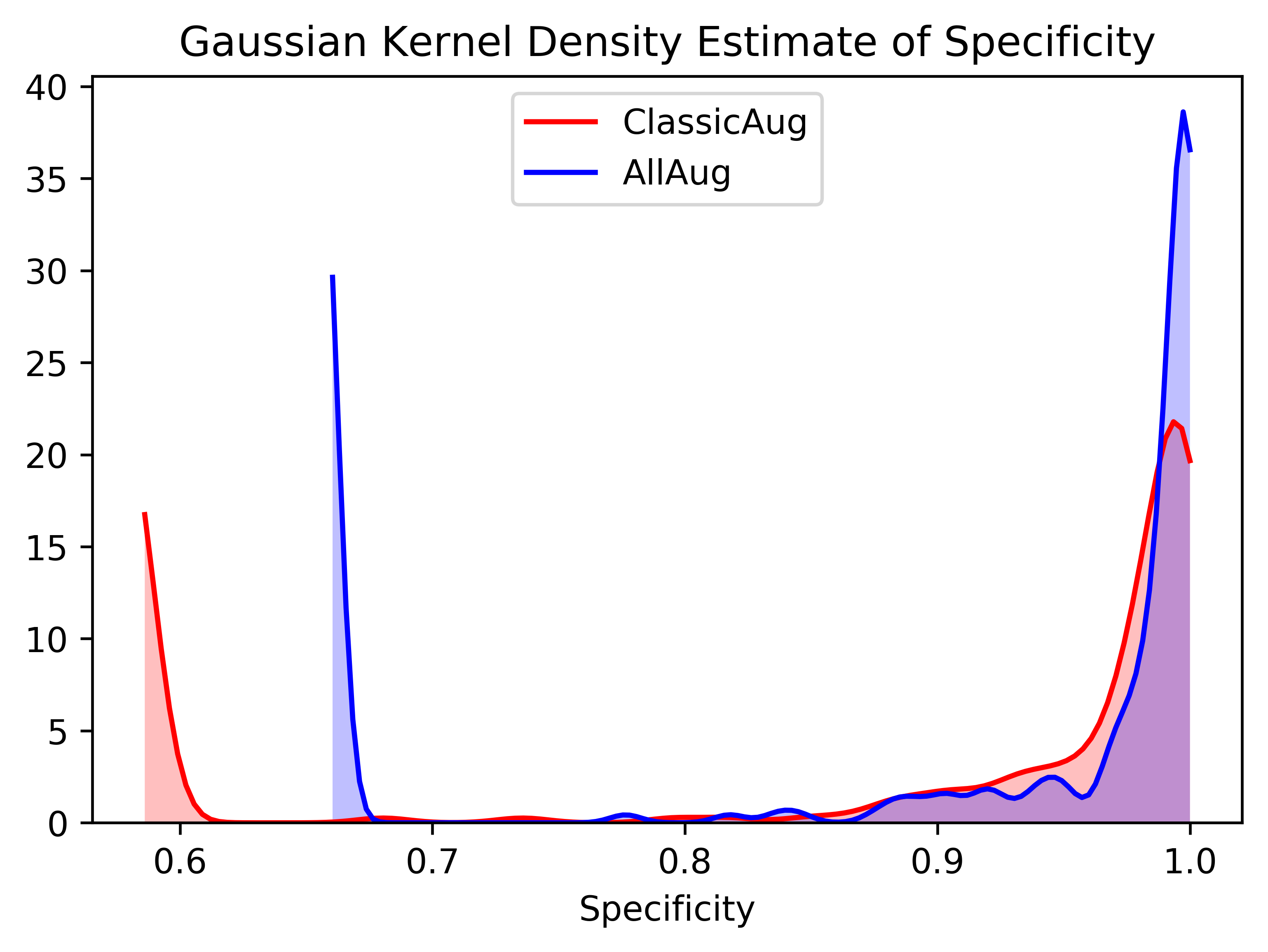}
  \caption{Evaluating the proposed method - comparing Dice coefficient (left), sensitivity (middle), and specificity (right) for ClassicAug and AllAug.}
  \label{fig:quant}
\end{figure*}

To further capture the segmentation performance improvement, we plot the Gaussian kernel density estimates of the Dice coefficient, the sensitivity, and the specificity obtained for the test images for the ClassicAug and AllAug (Figure~\ref{fig:quant}). The plots have been clipped to the range of values of the respective metrics and represent their probability density function estimates. The plots show higher peaks (which correspond to higher densities) at larger values of all the three metrics for AllAug as compared to ClassicAug. Moreover, the range of the specificity of AllAug is smaller than that of ClassicAug, meaning that combining classical augmentation with Mask2Lesion-based augmentation results in fewer mislabeled pixels.

\vspace{-2mm}
\section{Conclusion}    \label{sec:Conclusion}
In this work, we proposed Mask2Lesion, a conditional GAN-based model to generate skin lesion images from and constrained to binary masks, and used these newly generated images along with their corresponding masks to augment the training dataset for improving the segmentation accuracy of skin lesion images. In particular, we used the segmentation masks from the original dataset as input to the generative algorithm so as to avoid the manual annotation of the newly synthesized skin lesion images. We demonstrated that the generated lesion images are well-confined within the input mask boundaries, irrespective of the complexity of the masks. Our results showed a significant improvement in the segmentation accuracy when the training dataset for the segmentation network is augmented with these generated images. Future work direction will include extending this model to generate 3D medical images, such as CT and MR images.

\bibliographystyle{splncs04}
\bibliography{samplepaper.bib}




% LNCS Homepage, \url{http://www.springer.com/lncs}. Last accessed 4
% \end{thebibliography}
\end{document}